# The generalized Lindemann melting coefficient

Melvin M. Vopson[*,1], Nassina Rogers[1], Ian Hepburn[2]

[1*] *Faculty of Technology, School of Mathematics and Physics, University of Portsmouth, Portsmouth, UK*
[2] *Department of Space & Climate Physics, Faculty of Mathematics & Physical Sciences, University College London, UK*
*\* Correspondence to: melvin.vopson@port.ac.uk*

Lindemann developed the melting temperature theory over 100 years ago, known as the Lindemann criterion. Its main assumption is that melting occurs when the root-mean-square vibration amplitude of ions and atoms in crystals exceeds a critical fraction, $\eta$ of the inter-atomic spacing in crystals. The Lindemann coefficient $\eta$ is undefined and scientific papers report different $\eta$ values for different elements. Here we present previously unobserved data trends pointing to the fact that the Lindemann coefficient could be linked to the periodic groups of the periodic table, having an exact value for each element belonging to a given periodic group. We report 12 distinctive Lindemann coefficient values corresponding to 12 groups of the periodic table containing solid elements with identifiable melting temperature. Using these vales, the recalculation of the melting temperatures indicates a good match to the experimental values for 39 elements, corresponding to 12 out of 15 periodic groups. This newly observed result opens up the possibility of further refining the Lindemann melting criterion by stimulating analytical studies of the Lindemann coefficient in the light of this newly discovered result.

## 1. Introduction

The melting temperature of solids, also known as the melting point, is a material parameter very useful for technological applications, but also a matter of controversy as no detailed model explaining the melting temperature exists. Ideally, a working theory would allow a reliable prediction of the melting temperature in mono-atomic solids and complex alloys based on other known, or easy to estimate, physical parameters. Melting point of a solid is defined as the temperature at which a solid changes its state into a liquid at atmospheric pressure, so at the melting point the solid and liquid phases coexist in equilibrium. Although no analytical theory explaining the melting point exists, a widely used phenomenological approach to predict the melting temperature of solids was developed over 100 years ago by Lindemann, and it is known as the Lindemann criterion [1]. Lindemann approach is based on the observation that the amplitude of the thermal vibrations of ions / atoms in crystal increases with increasing the temperature of the crystal.

Lindemann postulated that melting occurs when the amplitudes of thermal vibrations are large enough for adjacent atoms / ions to partially occupy the same space. This could be viewed intuitively as crystals being stable and nearly static at low temperatures where the thermal vibrations are negligible, and solids shaking themselves to pieces at high temperatures. Gilvarry reformulated the criterion in terms of the mean-square amplitude of thermal vibrations, by stating that melting occurs when root-mean-square vibration amplitude exceeds a threshold value, which is typically taken as a fraction, $\eta$ of the inter-atomic spacing a in crystals [2]:

$$\sqrt{\langle u^2 \rangle} = \eta a \qquad (1)$$

$\eta$ is called the phenomenological Lindemann coefficient, initially assumed by Lindemann to be constant for all solids. In fact, the value of $\eta$ is not fixed and can take values in the range of 0.05 to 0.2 [3]. There have been many attempts to quantify this critical fraction using theoretical or phenomenological models applied to experimental data. Gupta and Sharma showed that the root-mean-square amplitude at the point of melting is around 10% of the interatomic distance, but the Lindemann coefficient values vary by a factor of three for different elements [4], including for elements of the same crystal structure. Correlations between various properties of solid-state materials, their melting temperature, Debye temperature and atomic mass numbers have been observed [5]. Overall, Lindemann melting criterion provides an effective way of quick estimating the melting temperatures of mono-atomic solids with ~ 20% accuracy. Its main criticism is that it is too simple and it only considers the solid phase. A true theory of melting would be able to comprehend both the solid and liquid phase. Several authors addressed this via thermodynamic approaches [6-9], but the predictions are still not accurate. In this article, we are revisiting the original melting temperature theory and we are analysing the existing experimental data in a different way, resulting in interesting trends and correlations previously not observed and unreported. These new results allowed us to generalize the values of the Lindemann coefficient and they offer a unique tool to further improve the theory of the melting temperature.

## 2. Theory

At high temperatures, quantum effects can be neglected, and the mean-square displacement $\langle u^2 \rangle$ of atoms is given by [10]:

$$\langle u^2 \rangle = \frac{9 k_B T}{4\pi^2 M \nu^2} \qquad (2)$$

where $M$ is the mass of the atom / ion, $k_B$ is the Boltzmann constant, $\nu$ is the vibration frequency and $T$ is the temperature. The Debye frequency, $\nu_D$ is defined as the highest allowed mode of vibration with all oscillators vibrating at the same frequency and phase in the crystal. It is reasonable to assume that, at the melting point of a solid, the ions / atoms have not only the highest allowed amplitude as dictated by the Lindemann criterion, but also the highest allowed frequency, Debye frequency, so that in $\nu = \nu_D$ in relation (2). Introducing the Debye temperature, $\theta_D$, as:

$$h\nu_D = k_B \theta_D \qquad (3)$$

where h is the Plank constant, then (2) becomes:

$$\langle u^2 \rangle = \frac{9 h^2 T}{4\pi^2 M k_B \theta_D^2} \qquad (4)$$

Relation (4) shows that the average vibration amplitude of atoms / ions in crystal increases linearly with the temperature. However, this cannot increase indefinitely, and Lindemann criterion offers the upper limit at the melting point, when the square-root of the average square vibration amplitude exceeds a certain fraction of the interatomic distance (see relation (1)). Introducing condition (1) in (4) at the melting point, then the general relation describing the melting temperature is:

$$T_m = \frac{4\pi^2 M k_B \eta^2 a^2}{9 h^2} \theta_D^2 \qquad (5)$$

Expressing the mass of the atom M in terms of the atomic mass number A (M = A / N$_A$), with N$_A$ being the well-known Avogadro constant, we obtain the following relationship between the melting temperature and the Debye temperature:

$$T_m = \frac{4\pi^2 A \eta^2 a^2 k_B}{9 N_A h^2} \theta_D^2 \qquad (6)$$

Equation (6) describes the melting temperature of a mono-atomic crystal, consisting of atoms with atomic mass number A, and its dependence on the square of the Debye temperature, $\theta_D^2$ and the square of the maximum allowed average vibration amplitude under Lindemann criterion, $<u^2> = \eta^2 a^2$. Relation (6) can be used to estimate the melting point of solids with considerable success. The main issues are related to the fact that the value of the Lindemann coefficient $\eta$ is not specified and the accuracy of the formula is in the range of 20%. By observing that relation (6) contains a number of constants, this can be further simplified as:

$$T_m = \xi \cdot A \eta^2 a^2 \theta_D^2 \tag{7}$$

Where $\xi$ is a constant given by: $\xi = \dfrac{4\pi^2 k_B}{9 N_A h^2} = 2.29 \times 10^{20} \left(\dfrac{mol}{JKs}\right)$. It is important to specify that the pseudo-Debye temperature $\theta_D$ can be defined at a given real temperature because the sound velocity in a crystal can vary with the temperature. However, most frequently utilized Debye temperatures in the literature are the low temperature at $0K$, $\theta_D(0K)$ and the high temperature taken at room temperature, $\theta_D(RT)$. It is reasonable to assume that the $\theta_D$ is in fact the $\theta_D(RT)$ throughout this study, as this is closer to the melting temperature.

## 3. Results

Figure 1 shows the melting temperature plotted as a function of the atomic mass number for 49 mono-atomic solids. The values of a, taken as the values of the nearest neighbour in the crystal, their room temperature Debye temperature, melting temperature, atomic mass A, and their crystal structure have been extracted from two data sources [11,12]. The data shows no predictable variations with points almost randomly scattered. Examining relation (7), we now define the ratio $\gamma = T_m / \theta_D^2$:

$$\gamma = \dfrac{T_m}{\theta_D^2} = \xi \cdot A \eta^2 a^2 \tag{8}$$

which shows that the ratio of the melting temperature to the square of the Debye temperature equals the product of the constant $\xi$, times the atomic mass number, the square of the interatomic spacing and the square of the Lindemann coefficient. According to (8), by plotting the ratio of the melting temperature to the square of the Debye temperature, $\gamma = T_m/\theta_D^2$, as a function of the atomic mass numbers, it is expected that this would display some kind of linear dependence on the atomic mass number. As expected, the graph of $T_m/\theta_D^2$ versus the atomic mass number, shown in figure 2, displays a consistent increase of the $T_m/\theta_D^2$ with A.

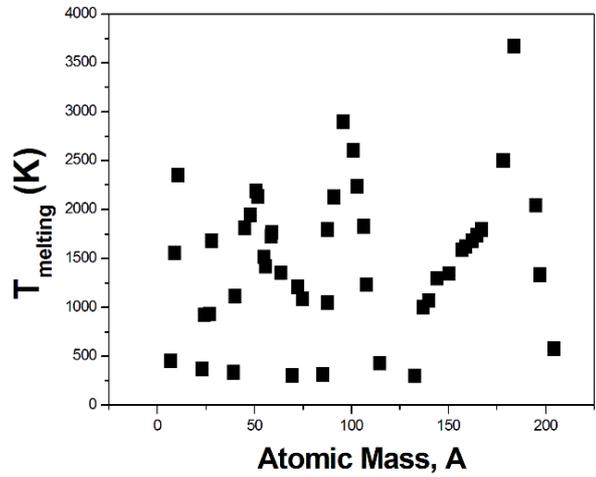

**Figure 1.** Melting temperature versus atomic mass number, A.

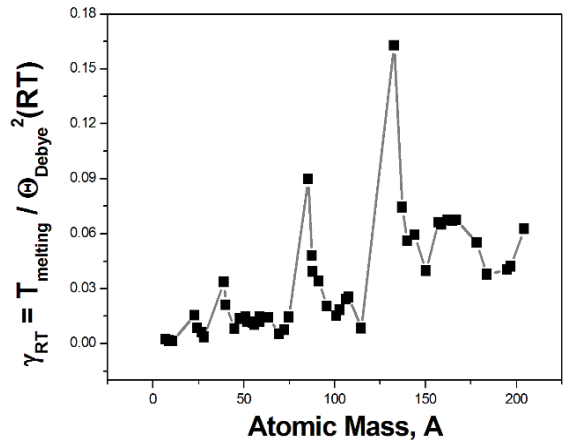

**Figure 2.** The ratio $\gamma_{RT}/\theta_D(RT) = T_m/\theta_D^2(RT)$ versus atomic mass number. A set of interesting peaks clearly emerged.

However, some unexplained interesting peak-like features occurring at specific atomic mass numbers with a given periodicity could be clearly identified in figure 2. At a closer inspection of the data, by identifying the elements corresponding to these peak features we were able to reveal a very interesting trend, in which the ratio $T_m/\theta_D^2$ of mono-atomic solids made up of elements in a given chemical group of the periodic table, are organized linearly as a function of their atomic mass numbers. For clarity, we show the same data from figure 2, re-plotted in figure 3, with chemical elements clearly marked on the graph. Besides identifying each relevant point on the peak-like features with its corresponding element in the periodic table, we also used a colour scheme to distinguish between the elements belonging to a given group in the periodic table, i.e. red for Group 1, blue for Group 2 and green for Group 3 (or Group 13 under the international naming convention).

Remarkably, it appears that the elements belonging to Group 1, Group 2 and Group 3 of the periodic table align linearly, as indicated in figure 3. The only exception is Thallium (Tl) from Group 3 that appears to be outside the trend. It is important to specify that the observed trends are applicable for solid elements and do not work for the radioactive elements, the heavy elements such as Lanthanides and Actinides and semiconductor elements. Although this is not fully understood, we believe it is related to the fact that these elements have more unstable nuclear and atomic structures, especially at higher temperatures. The data presented here relates to 49 solid chemical elements, mostly metallic. The results shown in figure 3 are intriguing because there is nothing in relation (8) that would suggest some kind of dependence of the ratio $T_m/\theta_D^2$ on the periodic groups of the periodic table, the Z number or the valence.

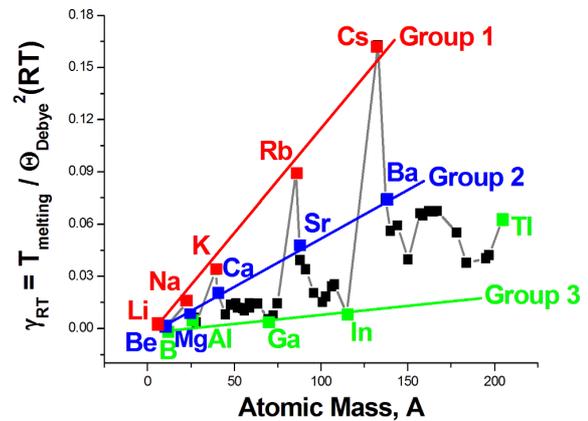

**Figure 3.** Data from figure 2 re-plotted here. The ratio $\gamma = T_m/\theta_D^2$ versus atomic mass number shows a clear linear correlation to elements belonging to groups in the periodic table.

Similar trends, not shown here, were also observed when the ratio of the melting point to the square of the low temperature Debye temperature was plotted, indicating that the dynamic of the melting process is dominated by the vibrational amplitudes of atoms / ions in crystals rather than their frequencies. These interesting observations prompted us to closely examine the dependence of the ratio $T_m/\theta_D^2$ for all the periodic groups of the periodic table containing solid elements. The results indicate that the observed linear relationship is in fact universal for all the groups of the periodic table containing solid elements.

Figure 4 shows the data for all 15 groups confirming this result. This allows an elegant method of extracting the Lindemann coefficient by properly processing the data. According to relation (8), plotting $T_m / \theta_D^2 \xi a^2$ versus A, results in a linear graph with the slope m, given by m = $\eta^2$, where

$$\xi = \frac{4\pi^2 k_B}{9 N_A h^2} = 2.29 \times 10^{20} \left(\frac{mol}{JKs}\right)$$

and a is the interatomic spacing taken as the nearest neighbour distance. Figure 5 shows the $T_m / \theta_D^2 \xi a^2$ versus A for 12 periodic groups, to which we applied a linear fit to the data. Out of 15 periodic groups, only 12 displayed the linear trend of $T_m / \theta_D^2 \xi a^2$ vs. A, or produced a reasonable linear fit, while groups 3, 13 and 14 were not included. From the linear fit, we were able to extract the slope m for each group set, followed by the Lindemann coefficient calculation:

$$\eta = \sqrt{m} \qquad (9)$$

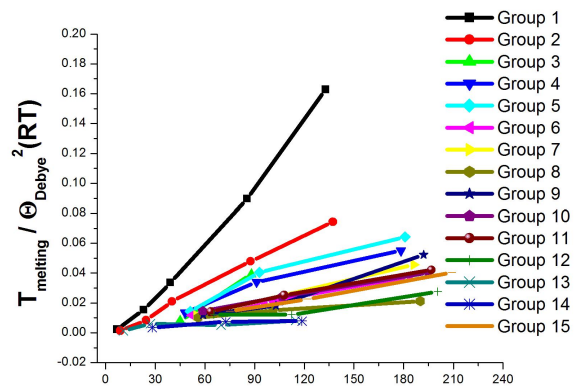

**Figure 4.** The ratio $\gamma_{RT} = T_m/\theta_D^2$ versus atomic mass number for mono-atomic crystals belonging to 15 groups of the periodic table. The data shows a clear linear correlation for elements belonging to any given group of the periodic table.

Our results indicate that we could allocate a single average Lindemann coefficient corresponding to all elements belonging to a given periodic group, which can be determined from the linear fit of the experimental data using (9). The results obtained in this work allowed the introduction of 12 distinctive Lindemann coefficients, applicable to all chemical elements within 12 periodic groups, with one Lindemann coefficient per periodic group.

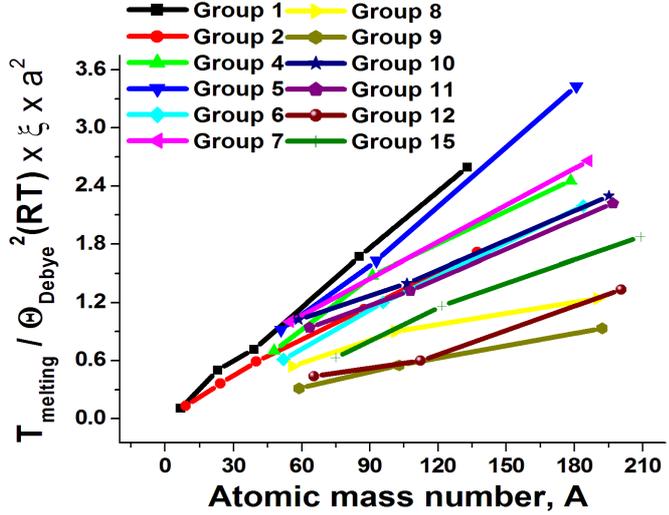

**Figure 5.** The ratio $T_m / \theta_D^2 \xi a^2$ versus atomic mass number for mono-atomic crystals belonging to 12 groups of the periodic table. A reasonable linear correlation for elements belonging to any given group of the periodic table is observed.

Although the linear fit produced a Lindemann coefficient for each periodic group, in order to verify our findings, the extracted values have been fed back into relation (7) and the theoretical melting temperature of each element has been calculated and compared with the experimental value. The results indicate a good agreement between the calculated melting temperature and the experimental values corresponding to 12 out of 15 periodic groups, while for the remaining 3 groups, although the linear dependence was observed, the extracted values of the Lindemann coefficient per group did not reproduce the correct melting temperatures for each element within the group.

Table 1 shows 39 chemical solid elements corresponding to the 12 periodic groups, together with their atomic mass number, atomic numbers, crystal structure, interatomic spacing, Debye temperatures at room temperature, the experimental melting temperatures and the calculated melting temperatures. Figure 6 shows a comparison between the melting temperature determined experimentally and calculated using our generalized Lindemann coefficients for 39 chemical mono-atomic solids corresponding to 12 periodic groups. Despite the large variations of the melting temperature with the atomic mass number, ranging from 234K in Hg to 3673K in W, the data indicates a good agreement between the experiment and the calculated values using our proposed Lindemann generalized coefficient values (see Fig. 6).

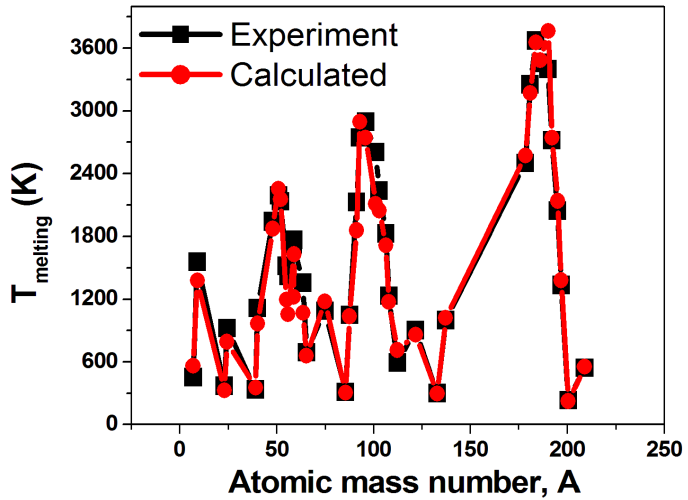

**Figure 6** Comparison between the experimental melting temperatures and the calculated values using our generalized Lindemann coefficients, for 39 elements of the periodic table.

**Table 1.** List of 39 chemical solid elements and their atomic mass number, crystal structure, nearest neighbour distance, Debye temperature, experimental melting temperature and calculated melting temperature.

| Element | A (g/mol) | Crystal | Nearest neighbour (pm) | $\theta_D$ (K) | $T_m$ (exp.) (K) | $T_m$ (calc.) (K) | $T_m$ (% difference) |
|---|---|---|---|---|---|---|---|
| Li | 6.94 | BCC | 302 | 448 | 453 | 562 | 24 |
| Be | 9.01 | HCP | 222 | 1031 | 1558 | 1380 | 12.8 |
| Na | 22.99 | BCC | 366 | 155 | 370 | 327 | 13.1 |
| Mg | 24.31 | HCP | 320 | 330 | 923 | 792 | 16.5 |
| K | 39.09 | BCC | 452.5 | 100 | 336 | 354 | 5.3 |
| Ca | 40.08 | FCC | 395 | 230 | 1115 | 967 | 15.3 |
| Ti | 47.9 | HCP | 289 | 380 | 1943 | 1873 | 3.7 |
| V | 50.94 | BCC | 262 | 390 | 2190 | 2252 | 5.6 |
| Cr | 51.99 | BCC | 291 | 424 | 2133 | 2153 | 1 |
| Mn | 54.94 | Other | 224 | 363 | 1517 | 1198 | 26.6 |
| Fe | 55.85 | BCC | 248 | 373 | 1422 | 1056 | 34.6 |
| Co | 58.93 | HCP | 250 | 386 | 1768 | 1631 | 8.4 |
| Ni | 58.69 | FCC | 249 | 345 | 1726 | 1222 | 41 |
| Cu | 63.55 | FCC | 256 | 310 | 1357 | 1069 | 27 |
| Zn | 65.38 | HCP | 350 | 237 | 692 | 659 | 5 |
| As | 74.92 | Other | 316 | 275 | 1089 | 1181 | 8.4 |
| Rb | 85.47 | BCC | 483.7 | 59 | 312 | 307 | 1.6 |
| Sr | 87.62 | FCC | 430 | 148 | 1050 | 1037 | 1.2 |
| Zr | 91.22 | HCP | 317 | 250 | 2127 | 1858 | 14.4 |
| Nb | 92.91 | BCC | 330 | 260 | 2745 | 2897 | 5.5 |
| Mo | 95.94 | BCC | 272 | 377 | 2895 | 2744 | 5.5 |
| Ru | 101.07 | HCP | 265 | 415 | 2606 | 2108 | 23.6 |
| Rh | 102.91 | FCC | 269 | 350 | 2236 | 2046 | 9.2 |
| Pd | 106.42 | FCC | 275 | 275 | 1828 | 1717 | 6.4 |
| Ag | 107.87 | FCC | 289 | 221 | 1234 | 1175 | 5 |
| Cd | 112.41 | HCP | 298 | 221 | 594 | 714 | 20 |
| Sb | 121.75 | Other | 291 | 200 | 903 | 861 | 4.8 |
| Cs | 132.91 | BCC | 523.5 | 43 | 301 | 297 | 1.3 |
| Ba | 137.33 | BCC | 435 | 116 | 1000 | 1022 | 2.2 |
| Hf | 178.49 | HCP | 313 | 213 | 2500 | 2572 | 2.8 |
| Ta | 180.95 | BCC | 286 | 225 | 3253 | 3172 | 2.5 |
| W | 183.85 | BCC | 274 | 312 | 3673 | 3655 | 0.5 |
| Re | 186.21 | HCP | 274 | 275 | 3458 | 3486 | 0.8 |
| Os | 190.2 | HCP | 268 | 400 | 3400 | 3764 | 10.7 |
| Ir | 192.22 | FCC | 271 | 228 | 2719 | 2742 | 0.8 |
| Pt | 195.08 | FCC | 277 | 225 | 2043 | 2138 | 4.6 |
| Au | 196.97 | FCC | 288 | 178 | 1336 | 1382 | 3.4 |
| Hg | 200.59 | Other | 301 | 92 | 234 | 225 | 4 |
| Bi | 208.98 | Other | 307 | 116 | 544 | 553 | 1.6 |

Previously reported results indicated a possible link between the type of crystal lattice of the elements and their melting temperatures [13]. We did observe some conclusive indirect links supporting this, but it appears that elements displaying multiple equilibrium crystallographic structures with very different melting temperatures are not obeying the generalization, i.e Carbon for example. Our results indicate that the melting temperature formula (7) is strongly sensitive to the interatomic spacing values, taken here as the nearest neighbour distance in the crystal. Small variations in these values lead to large changes in the melting point value. Since the measurement of interatomic spacing values are easily affected by temperature, pressure, crystal defects, impurities and instrument errors, large errors are expected to propagate when the generalized Lindemann coefficient and the melting temperatures are determined, which could explain why some data points show marginal deviations from the linear trends and some observed variations between the calculated and experimental melting temperatures. Moreover, the measurements of the Debye temperatures are also very imprecise, further contributing to enhancing the uncertainties in the melting temperature. If we assume that $\sigma_a$ and $\sigma_{\theta_D}$ are the uncertainties of the nearest neighbour spacing and the Debye temperature, then the overall uncertainty in the melting temperature is:

$$\frac{\sigma_{T_m}}{T_m} = 2\sqrt{\left(\frac{\sigma_a}{a}\right)^2 + \left(\frac{\sigma_{\theta_D}}{\theta_D}\right)^2} \tag{10}$$

To get a numerical idea of this, let us assume a conservative value of 5% relative uncertainty in the nearest neighbour and 10% in the Debye temperature, resulting in an estimated relative uncertainty of 22% for the melting temperature.

Figure 7 shows the periodic table of elements with the calculated Lindemann coefficients inserted at the top of each periodic group. Only periodic groups for which the obtained Lindemann values reproduced correctly the experimental values of the melting temperatures are highlighted in figure 7.

| 1 | 2 | 3 | 4 | 5 | 6 | 7 | 8 | 9 | 10 | 11 | 12 | 13 | 14 | 15 | 16 | 17 | 18 |
|---|---|---|---|---|---|---|---|---|---|---|---|---|---|---|---|---|---|
| $\eta_1$ 0.139 | $\eta_2$ 0.113 | - | $\eta_4$ 0.119 | $\eta_5$ 0.136 | $\eta_6$ 0.109 | $\eta_7$ 0.12 | $\eta_8$ 0.084 | $\eta_9$ 0.07 | $\eta_{10}$ 0.111 | $\eta_{11}$ 0.108 | $\eta_{12}$ 0.08 | - | - | $\eta_{15}$ 0.095 | - | - | - |
| 1 H 1 | | | | | | | | | | | | | | | | | 2 He 4 |
| 3 Li 7 | 4 Be 9 | | | | | | | | | | | 5 B 11 | 6 C 12 | 7 N 14 | 8 O 16 | 9 F 19 | 10 Ne 20 |
| 11 Na 23 | 12 Mg 24 | | | | | | | | | | | 13 Al 27 | 14 Si 28 | 15 P 31 | 16 S 32 | 17 Cl 35 | 18 Ar 40 |
| 19 K 39 | 20 Ca 40 | 21 Sc 45 | 22 Ti 48 | 23 V 51 | 24 Cr 52 | 25 Mn 55 | 26 Fe 56 | 27 Co 59 | 28 Ni 59 | 29 Cu 64 | 30 Zn 65 | 31 Ga 70 | 32 Ge 73 | 33 As 75 | 34 Se 79 | 35 Br 80 | 36 Kr 84 |
| 37 Rb 85 | 38 Sr 88 | 39 Y 89 | 40 Zr 91 | 41 Nb 93 | 42 Mo 96 | 43 Tc (98) | 44 Ru 101 | 45 Rh 103 | 46 Pd 106 | 47 Ag 108 | 48 Cd 112 | 49 In 115 | 50 Sn 119 | 51 Sb 122 | 52 Te 128 | 53 I 127 | 54 Xe 131 |
| 55 Cs 133 | 56 Ba 137 | 57 *La 139 | 72 Hf 178 | 73 Ta 181 | 74 W 184 | 75 Re 186 | 76 Os 190 | 77 Ir 192 | 78 Pt 195 | 79 Au 197 | 80 Hg 201 | 81 Tl 204 | 82 Pb 207 | 83 Bi 209 | 84 Po (209) | 85 At (210) | 86 Rn (222) |
| 87 Fr (223) | 88 Ra (226) | 89 †Ac (227) | 104 Rf (267) | 105 Db (268) | 106 Sg (271) | 107 Bh (272) | 108 Hs (270) | 109 Mt (276) | 110 Ds (281) | 111 Rg (280) | 112 Cn (285) | 113 Nh (284) | 114 Fl (289) | 115 Mc (288) | 116 Lv (293) | 117 Ts (294) | 118 Og (294) |

**Figure 7.** Periodic table of elements and the corresponding Lindemann melting coefficients of each group containing solids. Only the highlighted elements were included in the analysis.

## 4. Conclusions

By analysing the experimental data of 49 chemical solid elements we determined an interesting relationship between the ratio of the melting temperature to the square of the Debye temperature and the corresponding atomic mass number, which shows a linear dependence with elements belonging to the same period group of the periodic table, aligned on the same linear graph. These previously unobserved data trends allowed us to demonstrate that the Lindemann melting coefficient $\eta$ is in fact an exact value for each element belonging to a given periodic group of the periodic table, resulting in 15 distinctive Lindemann coefficients, applicable to all chemical elements, with one Lindemann coefficient per periodic group. When testing the results obtained here against the experimental melting temperature values for all 49 elements, we obtained good agreement for only 39 of them belonging to 12 periodic groups. The observed generalization works for solid metallic elements, but it is not applicable to the radioactive elements, the heavy elements such as Lanthanides and Actinides, and semiconductor elements.

A refined theoretical Lindemann model of melting must account for the observed relationship, as well as the interactions taking place in more complex solids, and this could lead to further improvements in the theory of melting temperatures applicable possibly not only to mono-atomic solids, but also to more complex alloys and chemicals. At a closer examination of relation (8), it appears that the observed trends can only be accommodated by parameters hidden within the phenomenological Lindemann parameter, η. Combining the findings of this work with a first principle computational approach [14], as well as with other previous studies in which a generalization of Lindemann melting criterion was proposed for 2D materials [15,16], could open up the possibility of further refining the Lindemann melting criterion by identifying the possible hidden parameters within Lindemann coefficient and its dependencies on the periodic groups.